\documentclass{scrartcl}

\usepackage{authblk}
\usepackage[utf8]{inputenc}
\usepackage{hyperref}
\usepackage{graphicx}
\usepackage{hyperref}
\usepackage{multirow}

\begin{document}

\title{Testing performance with and without Block Low Rank Compression in MUMPS and the new PaStiX 6.0 for JOREK nonlinear MHD simulations}
\author[1]{Richard Nies}
\author[1]{Matthias Hoelzl}
\affil[1]{Max-Planck-Institute for Plasma Physics, Boltzmannstr. 2, 85748 Garching b. M., Germany}
\date{\today}
\maketitle
\begin{abstract}
The interface to the MUMPS solver was updated in the JOREK MHD code to support Block Low Rank (BLR) compression and an interface to the new PaStiX solver version 6 has been implemented supporting BLR as well. First tests were carried out with JOREK, which solves a large sparse matrix system iteratively in each time step. For the preconditioning, a direct solver is applied in the code to sub-matrices, and at this point BLR was applied with the results being summarized in this report. For a simple case with a linearly growing mode, results with both solvers look promising with a considerable reduction of the memory consumption by several ten percent was obtained. A direct increase in performance was seen in particular configurations already.

The choice of the BLR accuracy parameter $\epsilon$ proves to be critical in this simple test and also in more realistic simulations, which were carried out only with MUMPS due to the limited time available. The more realistic test showed an increase in run time when using BLR, which was mitigated when using larger values of $\epsilon$. However, the GMRes iterative solver does not reach convergence anymore when $\epsilon$ is too large, since the preconditioner becomes too inaccurate in that case. It is thus critical to use an $\epsilon$ as large as possible, while still reaching convergence. More tests regarding this optimum will be necessary in the future. BLR can also lead to an indirect speed-up in particular cases, when the simulation can be run on a smaller number of compute nodes due to the reduced memory consumption.

\end{abstract}

\newpage
\tableofcontents

\section{Background}
\subsection{JOREK solver}
The \verb|JOREK| \cite{Huysmans_2007} solver always involves LU factorisations, either to solve the entire matrix directly (generally used when the simulation is run axisymmetrically) or as a preconditioner before using the iterative solver GMRes. We focus on the latter case. 

In the preconditioner, the blocks of toroidal harmonics are assumed to be decoupled from one another (i.e.\ off-diagonal blocks are not considered in the preconditioner), which greatly reduces the memory requirements and the runtime. This approach comes to its limits when the system is very nonlinear, as the preconditioner will be inaccurate and many GMRes iterations may be needed to converge to the solution.

\subsection{LU factorisation of sparse matrix}
The blocks of harmonics are sparse matrices (i.e.\ a small percentage of entries is non-zero) and naive algorithms used for dense matrices should thus not be applied here. Indeed, these would lead to a substantial fill-in: the combined storage cost for the $L$ and $U$ factors would greatly exceed that of the original matrix. The \verb|JOREK| code makes use of either \verb|PaStiX| \cite{henon:inria-00346017} or \verb|MUMPS| \cite{doi:10.1137/S0895479899358194}, both of which are able to efficiently handle sparse matrices.

In the following, we will briefly present the main steps of these solvers, without going into the details. More information can be found e.g. in \cite{mary:tel-01929478}.

In the \textbf{analysis} phase, a graph ordering tool is called, generally \verb|SCOTCH| or \verb|METIS|. These represent the matrix as a graph and an elimination tree is constructed to reduce fill-in. This tree is traversed as each node is eliminated and its contribution to nodes further up the tree are  computed (corresponding to fill-in). This corresponds to an update of nodes up the tree which can be done either immediately after elimination (fan-out or right-looking), as late as possible before the elimination of the node further up the tree (fan-in or left-looking) or it can be carried up the tree, accumulating further updates from other nodes in so called fronts (fan-out or right-looking)\footnote{
PaStiX is using a fan-in strategy in the communication. In the sense that  $A\cdot B$ are accumulated locally on the origin node, and then sent to the remote node when all contributions are applied, such that the remote node can perform the operation $C=C+sum(A\cdot B)$. Except for this specific case in distributed, all the algorithm is fan-out/right-looking oriented to get more parallelism, even for the accumulation of the remote contribution. MUMPS, in contrast, based on the multi-frontal approach, can be considered as fan-in/left-looking oriented algorithm.}.

In the \textbf{factorisation} phase, the factors of the $L$ and $U$ matrices are computed, following the elimination tree obtained in the analysis phase. This is typically the most CPU intensive step, although the factorisation does not need to be repeated for every time step in a typical \verb|JOREK| simulation, since the preconditioning matrix is reused as long as reasonable convergence is obtained in GMRes.

In the \textbf{solve} phase, the solution is computed using standard forward and back-substitution methods. A solve also has to be repeated for every iteration in GMRes.

\subsection{BLR compression}

Even with the use of graph ordering tools, the fill-in can be substantial and the required memory storage for the factors of $L$ and $U$ can be very high. As a possible solution to this problem, the factors could be compressed using Block-Low-Rank (BLR) compression.

Let $A$ be a matrix of size $m\times n$. Let $k_\epsilon$ be the approximated numerical rank of $A$ at accuracy $\epsilon$. $A$ is a low-rank matrix if there exist three matrices $U$ of size $m\times k_\epsilon$ , $V$ of size $n\times k_\epsilon$ and $E$ of size $m\times n$ such that :
\begin{equation}
    A= U \cdot V^T +E,
\end{equation}
where $||E||_2 \leq \epsilon$ and $k_\epsilon \cdot (m + n) < mn$. The last condition implies lower dimensionality and thus also lower storage costs for combined U and V than for original A, provided the matrix is sufficiently dense. The basic idea is to represent the matrix as the product of two skinny matrices. Note that lossless compression is possible as the accuracy $\epsilon$ can be set to $0$.

Besides possible memory gains, using BLR compression can in principle also reduce the runtime, as basic matrix operations are accelerated. Indeed, for $m=n$, matrix-matrix products go from $2n^3$ (for dense matrices) to $2kn(n+2k)$ operations, while a triangular solve goes from $n^3$ to $3kn^2$ operations. The complexity is thus reduced for $k<n/2$ and $k<n/3$ respectively.

The $L$ and $U$ matrices are generally not low-rank, but individual sub-matrices may be efficiently compressed. There is no general admissibility criterion to determine if a block should be stored in low-rank. This depends on the solver used, see \cite{amestoy:hal-01505070} (\verb|MUMPS|) and \cite{pichon:hal-01824275} (\verb|PaStiX|) for more details on the implementation of BLR compression in these solvers.

\subsubsection{Runtime performance in JOREK}
\label{sec_runtime_perf}

The runtime of the \verb|JOREK| solver will be impacted by the use of BLR compression. Before testing this in detail in the rest of this report, we layout how the performance is affected by BLR compression.

The \textbf{analysis} step takes a longer time due to the additional search for suitable matrices to compress. However, the impact on the total runtime is negligible, as the analysis only has to be done once at the simulation (re-)start.

Two opposing effects will play a role in the \textbf{factorisation} step: the compression of the LU blocks leads to an overhead but the compressed matrices lead to faster basic matrix operations (see above). In practice, the time for factorisation seems to be generally increased with BLR in our tests, although higher values of $\epsilon$ can reduce or even revert this effect.

The same conclusions hold for the \textbf{solve} step, as the (de-)compression induces an overhead, which the faster matrix operations can cancel\footnote{This overhead can be avoided and will be avoided in both PaStiX and MUMPS in the future.}

The runtime performance of the \textbf{GMRes} iterative solver is set by the number of iterations needed to reach convergence and the time needed for a basic solve (see step above), as it is repeated for every GMRes iteration. As for the factorisation and solve steps above, this means that the use of BLR causes an overhead but larger values of $\epsilon$ will not always be beneficial here. Indeed, they may reduce the number of matrix operations and thereby reduce runtime, but a too large value of $\epsilon$ deteriorates the accuracy of the preconditioner and leads to worse convergence in GMRes, increasing the number of iterations necessary to reach convergence. Taken to the extreme, this can stop the simulation, as the GMRes iterative solver is given a maximum number of steps to reach convergence in \verb|JOREK|.

\section{BLR compression in MUMPS}
\subsection{Resolution scan}
\label{sec_MUMPS_resol}
\subsubsection{Basic setup}
All simulations in this report were carried out on a Linux cluster, where each compute node is equipped with 2x Intel Xeon Gold 6130 CPUs with 16 cores and 2.1 GHz base clock speed, AVX 512, ``Skylake'' architecture, 22 MB L3 cache, and fast Omnipath interconnect.

We first perform basic resistive ballooning mode simulations (a test case called \verb|inxflow| in \verb|JOREK|) with 2 toroidal harmonics ($n=0,6$) and 4 toroidal planes. The simulations are restarted during the linear growth phase for 10 time steps, with the factorisation enforced to be carried out in every time step. The iterative solver GMRes is given a tolerance of $10^{-7}$ and $50$ maximal iterations to reach convergence. Each simulation is run on one compute node with 2 MPI tasks, with 2 OpenMP threads each. 

The resolution was varied from \verb|(n_flux, n_tht) = (16, 20)| up to \\\verb|(n_flux, n_tht) = (128, 160)|, where \verb|n_flux| denotes the radial and \verb|n_tht| the poloidal number of grid points, in 6 steps of approximate factors $(\sqrt{2}, \sqrt{2})$. For each resolution, the runtime performance and memory consumption of the solver MUMPS are investigated, with and without BLR compression, and with different values of the BLR tolerance: $\epsilon = 0, 10^{-16}, 10^{-12}, 10^{-8}, 10^{-4}$.

\subsubsection{Memory consumption}

\begin{figure}[!ht]
	\includegraphics[scale=1]{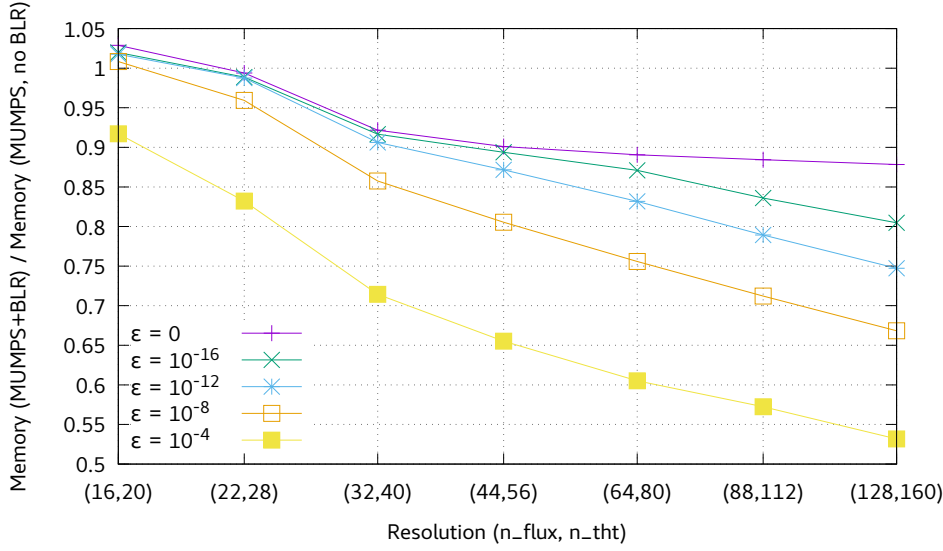}
	\caption{Memory gains from BLR compression}
	\label{fig_resol_memory}
\end{figure}

The total memory consumption in MB of the MUMPS solver depending on resolution is shown in Tab~\ref{tab_memory} and Fig.~\ref{fig_resol_memory}. Note that the blocks corresponding to nonzero toroidal harmonics have twice the dimensionality of the $n=0$ block, such that $1/5$ of the total memory indicated is used in the $n=0$ block and $4/5$ in the $n=6$ block.

Significant memory gains can be made using BLR compression. These gains are exacerbated for large values of $\epsilon$ (as the matrices can be compressed more effectively) and for large problem sizes (as there are more opportunities for compression). 

\begin{table}[!ht]
\begin{tabular}{c|c c c c c c}%
\verb|n_flux,n_tht| &  No BLR &  $\epsilon=0$ &  $\epsilon=10^{-16}$ &  $\epsilon=10^{-12}$ &  $\epsilon=10^{-8}$ &  $\epsilon=10^{-4}$ \\[1pt]\hline
$16,20$ & $  799$ & $  822$ & $  815$ & $  813$ & $  806$ & $  733$\\
$22,28$ & $  1762$ & $  1751$ & $  1742$ & $  1740$ & $  1690$ & $  1466$\\
$32,40$ & $  4365$ & $  4023$ & $  4001$ & $  3956$ & $  3743$ & $  3118$\\
$44,56$ & $  9047$ & $  8151$ & $  8086$ & $  7886$ & $  7286$ & $  5928$\\
$64,80$ & $  20891$ & $  18603$ & $  18193$ & $  17375$ & $  15787$ & $  12644$\\
$88,112$ & $  45628$ & $  40353$ & $  38148$ & $  36015$ & $  32494$ & $  26119$\\
$128,160$ & $  101035$ & $  88741$ & $  81299$ & $  75508$ & $  67516$ & $  53717$
\end{tabular}
\caption{Memory consumption in the MUMPS resolution scan in Megabytes}
\label{tab_memory}
\end{table}

Note, however, that large values of $\epsilon$ can significantly deteriorate the quality of the preconditioning, leading to an increased number of iterations for GMRes (see Sec.~\ref{sec_runtime_perf}). Depending on the required tolerance for GMRes and the number of iterations, this can even prevent convergence, which was the case in the highest resolution simulations with $\epsilon = 10^{-4}$.

\subsubsection{Runtime}

The runtimes listed below are averages over the simulations' 10 time steps. The activation of the Block-Low-Rank feature leads to an increase in the analysis phase's runtime by a factor of approximately $2$. It was however left out in the following, as the analysis only has to be performed once and its impact on the performance is thus negligible.

\begin{table}[!ht]
\begin{tabular}{c|c c c c c c}%
\verb|n_flux,n_tht| &  No BLR &  $\epsilon=0$ &  $\epsilon=10^{-16}$ &  $\epsilon=10^{-12}$ &  $\epsilon=10^{-8}$ &  $\epsilon=10^{-4}$ \\[1pt]\hline
$16,20$    & $   1.1$ & $   1.5$ & $   1.5$ & $   1.4$ & $   1.4$ & $   1.2$\\
$22,28$    & $   2.6$ & $   3.7$ & $   3.7$ & $   3.6$ & $   3.5$ & $   2.7$\\
$32,40$    & $   7.2$ & $  10.3$ & $  10.0$ & $  10.1$ & $   9.0$ & $   7.0$\\
$44,56$    & $  16.6$ & $  23.5$ & $  23.7$ & $  22.5$ & $  19.3$ & $  14.2$\\
$64,80$    & $  44.1$ & $  64.6$ & $  62.5$ & $  56.5$ & $  46.1$ & $  32.9$\\
$88,112$   & $ 119.5$ & $ 168.6$ & $ 144.6$ & $ 126.6$ & $ 101.9$ & $  74.1$\\
$128,160$  & $ 330.1$ & $ 496.0$ & $ 388.6$ & $ 319.6$ & $ 253.4$ & $ 164.9$
\end{tabular}
\caption{Runtime for factorisation step (in s)}
\label{tab_resol_facto}
\end{table}

\begin{table}[!ht]
\begin{tabular}{c|c c c c c c}%
\verb|n_flux,n_tht| &  No BLR &  $\epsilon=0$ &  $\epsilon=10^{-16}$ &  $\epsilon=10^{-12}$ &  $\epsilon=10^{-8}$ &  $\epsilon=10^{-4}$ \\[1pt]\hline
$16,20$    & $   2.0$ & $   2.7$ & $   2.7$ & $   2.6$ & $   2.6$ & $   2.5$\\
$22,28$    & $   4.5$ & $   6.1$ & $   6.1$ & $   6.1$ & $   5.8$ & $   5.3$\\
$32,40$    & $  11.1$ & $  15.7$ & $  15.2$ & $  15.5$ & $  14.2$ & $  13.2$\\
$44,56$    & $  24.4$ & $  34.0$ & $  34.3$ & $  33.3$ & $  29.6$ & $  25.8$\\
$64,80$    & $  59.9$ & $  88.2$ & $  86.1$ & $  80.1$ & $  84.0$ & $  \infty$\\
$88,112$   & $ 151.7$ & $ 217.1$ & $ 191.6$ & $ 175.0$ & $ 151.0$ & $  \infty$\\
$128,160$  & $ 410.8$ & $ 626.2$ & $ 499.2$ & $ 442.7$ & $ 428.8$ & $  \infty$
\end{tabular}
\caption{Runtime for entire time step (in s)}
\label{tab_resol_iter}
\end{table}

\begin{figure}[!ht]
	\includegraphics[scale=1]{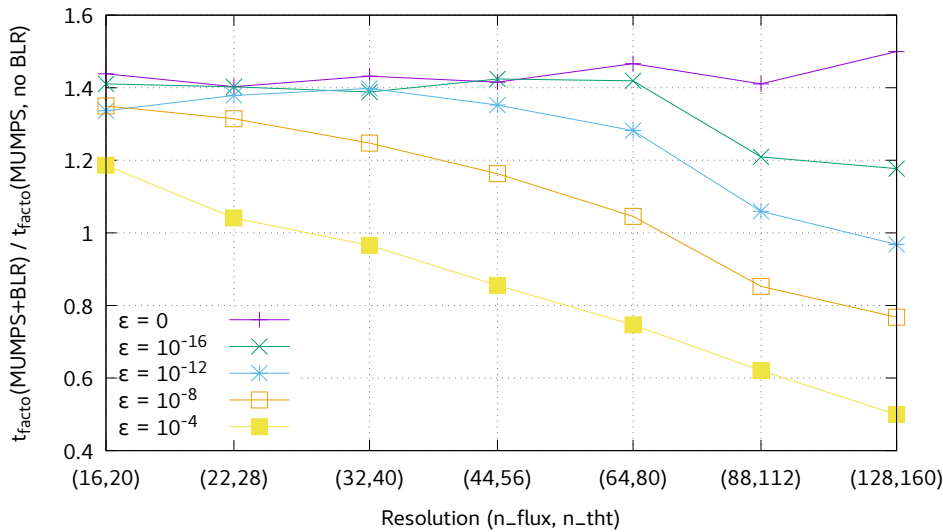}
	\caption{Runtime for factorisation step}
	\label{fig_resol_runtime_facto}
\end{figure}

The runtime for the factorisation step of the MUMPS solver is shown in Tab.~\ref{tab_resol_facto} and Fig.~\ref{fig_resol_runtime_facto}. Here, the use of BLR compression can be detrimental (overhead caused by the compression of LU factors) or beneficial (speedup of matrix operations). Which of these two effects dominates depends on the size of the problem and the choice of $\epsilon$.

\begin{figure}[!ht]
	\includegraphics[scale=1]{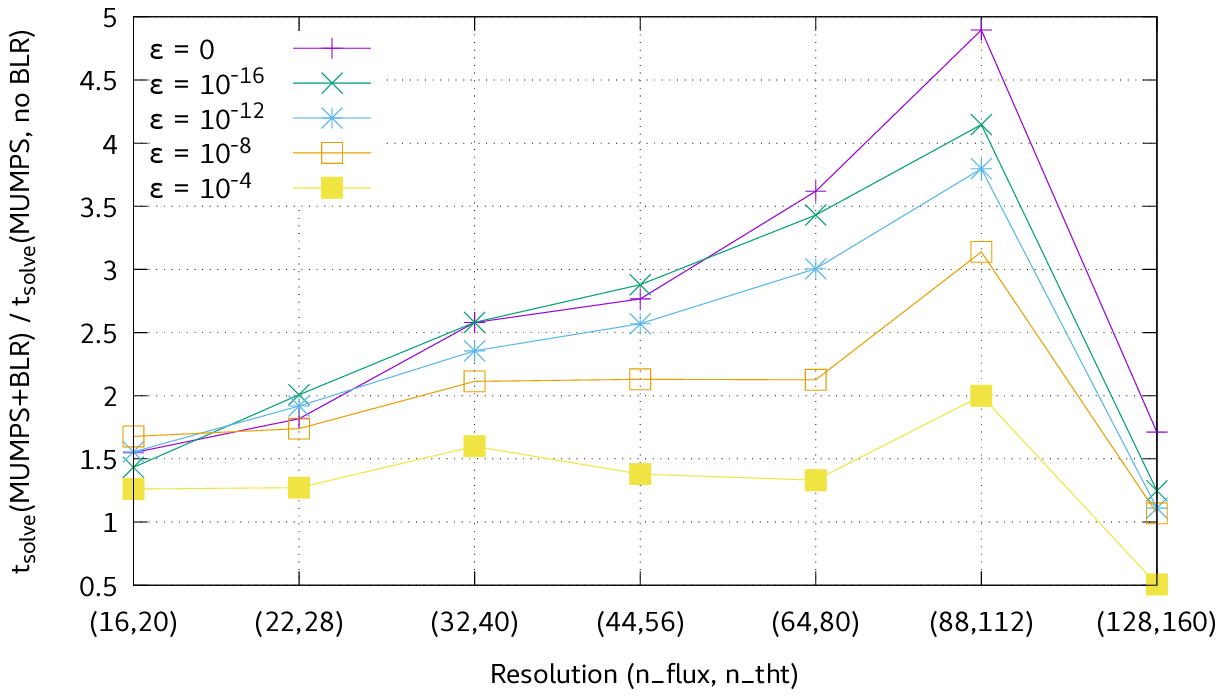}
	\caption{Runtime for solution step}
	\label{fig_resol_runtime_solve}
\end{figure}

The runtime for the solution step of the MUMPS solver is shown in Fig.~\ref{fig_resol_runtime_solve}. The solution step generally incurs an overhead from compression, which can however be mitigated at high resolutions, where the decrease in the number of operations can compensate for the overhead.

\begin{figure}[!ht]
	\includegraphics[scale=1]{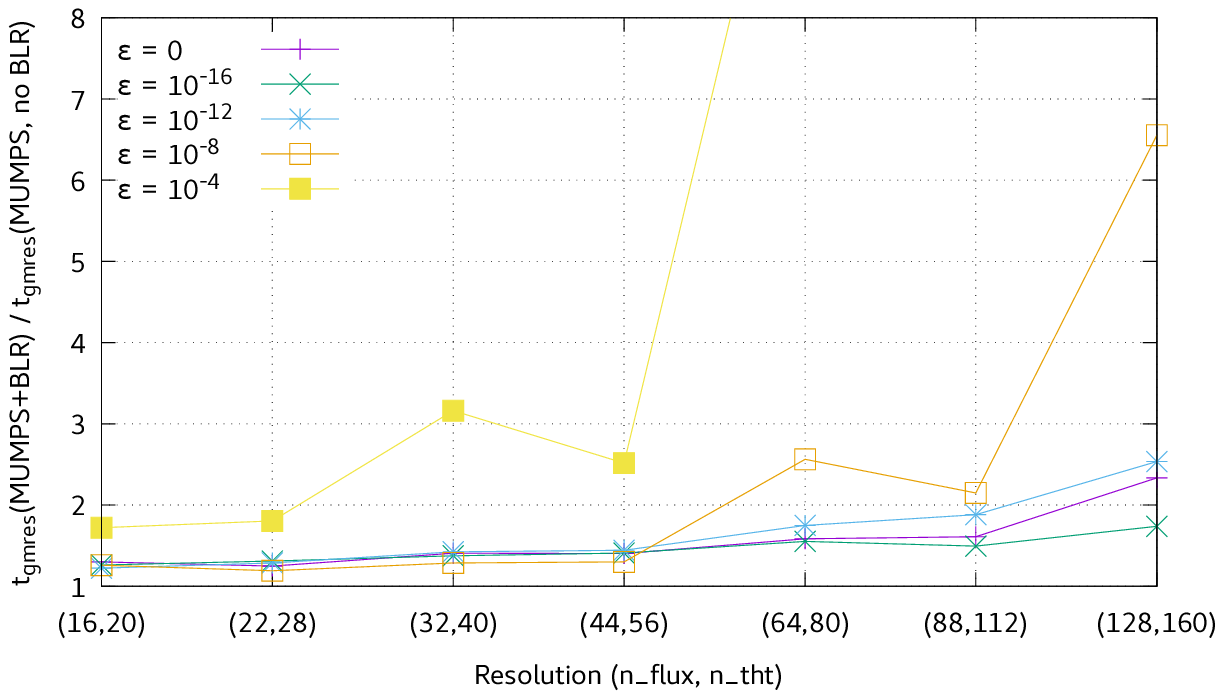}
	\caption{Runtime for GMRes}
	\label{fig_resol_runtime_gmres}
\end{figure}

The runtime for the GMRes step of the JOREK solver is shown in Fig.~\ref{fig_resol_runtime_gmres}, where the strong spike for $\epsilon = 10^{-4}$ reflects the loss of convergence. A slowdown is observed at all resolutions as the same overhead mentioned in the solution step comes into play here, as well as larger number of iterations in the GMRes solver when $\epsilon$ is too high and the preconditioner is too inaccurate.

The reduction of the time for the solution phase in Fig.~\ref{fig_resol_runtime_solve} at the highest resolution leaves open the possibility that this phase is actually accelerated at even larger resolutions. This could then also possibly lead to a reduction in the time for GMRes, if it was possible to compensate for both the (de-)compression overhead and the higher number of iterations caused by the preconditioner's inaccuracy.

\begin{figure}[!ht]
	\includegraphics[scale=1]{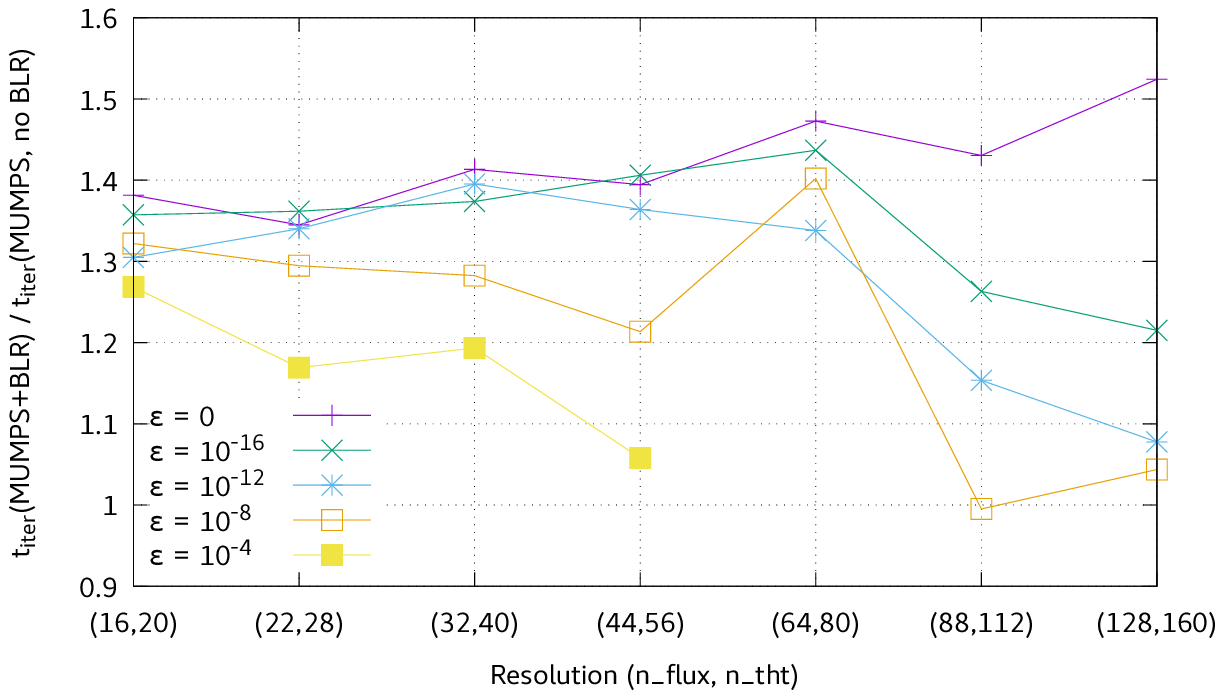}
	\caption{Average runtime per time step}
	\label{fig_resol_runtime_iter}
\end{figure}

The average total runtime per time step shown in Tab.~\ref{tab_resol_iter} confirms the generally increased runtime when using BLR compression. For large values of $\epsilon$, this increase is quite modest. In our simulations at highest resolution, the optimal choice is $\epsilon = 10^{-8}$, where the increase in runtime is minimal (no BLR: $411$ s, $\epsilon = 10^{-8}$: $429$ s) while the memory requirements are substantially lower (no BLR: $101$ GB, $\epsilon = 10^{-8}$: $68$ GB). The high performance observed here is due to the reduced runtime for factorisation, largely offsetting the increase in runtime from GMRes. The peculiar spike in the runtime for $\epsilon = 10^{-8}$ and a resolution of \verb|(n_flux, n_tht) = (64, 80)| is due to an increase in the number of iterations of GMRes for $3$ out of the $10$ time steps, suggesting that fluctuations in the preconditioner's accuracy can come into play for larger values of $\epsilon$.

However, the \verb|JOREK| solver generally keeps the preconditioning LU factors for several timesteps, until it does not satisfactorily precondition the matrix anymore, i.e. the number of GMRes iterations in the previous time step has passed a certain threshold. As a consequence, for many time steps only the solve step and GMRes have to be performed. For too large $\epsilon$, this could increase the runtime substantially, as GMRes is repeated several times for each LU factorisation. Additionally, the LU factorisation might need to be repeated after a smaller number of time steps. However, the runtimes for the solution phase and GMRes phases might be reduced at very high resolutions due to the reduced number of operations, as Figs.~\ref{fig_resol_runtime_solve} and \ref{fig_resol_runtime_gmres} seem to suggest.

The accuracy $\epsilon$ should be chosen as large as possible to allow for significant memory gains and keep the runtime as low as possible, without being too high, such that GMRes convergence is not too greatly impaired. This choice of $\epsilon$ will depend on the exact problem at hand as well as the setup of the GMRes solver.

\subsection{More realistic simulation}
\label{sec_spi}
The goal of this section is to ascertain the benefits of BLR compression during the nonlinear phase, where the preconditioning matrix is not as effective and the convergence of GMRes is thus even more critical, as well as its general usefulness in a typical \verb|JOREK| simulation.

We thus turn to simulations of shattered pellet injections (SPI), running for a larger number of time steps ($7330$) through linear and nonlinear phases, with 6 toroidal harmonics ($n=0...5$) and 32 toroidal planes. The resolution of all simulations is \verb|(n_flux,n_tht)| = (56, 138). The maximal number of iterations of the GMRes iterative solver was set to $100$ while its tolerance was set to $10^{-6}$ and $10^{-7}$, to investigate its effects on the convergence properties when using compression. Indeed, we force the recalculation of the preconditioner (i.e. a factorisation) every time the number of steps in GMRes exceeds $20$, such that a lower GMRes tolerance will lead to more factorisations. 

We again compare simulations with different MUMPS setups: without BLR and with BLR, using different values of $\epsilon$. For the higher GMRes tolerance value of $10^{-6}$, the simulations with $\epsilon = 10^{-8}, 10^{-4}$ did not yield satisfactory convergence in GMRes, the former in the linear and the latter in the nonlinear phase. We thus examine only the values $\epsilon = 0, 10^{-16}, 10^{-12}, 10^{-10}$.

The total memory consumption of the MUMPS solver in GB amounts to 
\begin{itemize}
	\item No BLR: \hspace{30pt}       $166.0$ ($100\%$)
	\item BLR, $\epsilon = 0$: \qquad $147.1$ ($88.6\%$)
	\item BLR, $\epsilon = 10^{-16}$: $139.8$ ($84.2\%$)
	\item BLR, $\epsilon = 10^{-12}$: $132.6$ ($79.9\%$)
	\item BLR, $\epsilon = 10^{-10}$: $126.8$ ($76.4\%$)
\end{itemize}

The runtime performance data of all simulations is given in Tab.~\ref{tab_data_spi_runs}. Note that the runtime for the factorisation and solve steps is independent of the GMRes tolerance value.

\begin{table}[h!]
\centering
\small
\begin{tabular}{l l | c c c c c}
GMRes tol. &  & No BLR &  $\epsilon=0$ &  $\epsilon=10^{-16}$ &  $\epsilon=10^{-12}$ &  $\epsilon=10^{-10}$\\[1pt]\hline
           & Nbr. of factorisations   & $786$   & $716$   & $950$   & $805$  & $816$  \\
$10^{-6}$  & Avg. time per iter. (s)  & $47.0$  & $65.7$  & $68.2$  & $63.3$ & $61.9$ \\      
           & Avg. nbr. of GMRes iter. & $14.5$  & $14.5$  & $15.6$  & $14.7$ & $14.7$ \\
           & Avg. time per GMRes (s)  & $13.5$  & $28.8$  & $28.9$  & $26.4$ & $25.0$ \\[1pt]\hline
           & Nbr. of factorisations   & $3145$  & $3033$  & $3044$  & $3026$ &  \\
$10^{-7}$  & Avg. time per iter. (s)  & $76.6$  & $109.4$ & $104.9$ & $99.7$ &  \\ 
           & Avg. nbr. of GMRes iter. & $21.2$  & $21.0$  & $21.3$  & $20.8$ &  \\     
           & Avg. time per GMRes (s)  & $18.8$  & $40.9$  & $38.8$  & $35.8$ &  \\[1pt]\hline
           & Avg. time per facto. (s) & $50.7$  & $67.9$  & $63.1$  & $58.3$ & $55.2$ \\
           & Avg. time per solve (s)  & $0.46$  & $1.78$  & $1.63$  & $1.58$ & $1.45$   
\end{tabular}
\caption{Performance of BLR compression in SPI simulations}
\label{tab_data_spi_runs}
\end{table}

\subsubsection{Runtime performance for a GMRes tolerance of $10^{-6}$}

\begin{figure}[ht!]
	\includegraphics[scale=1]{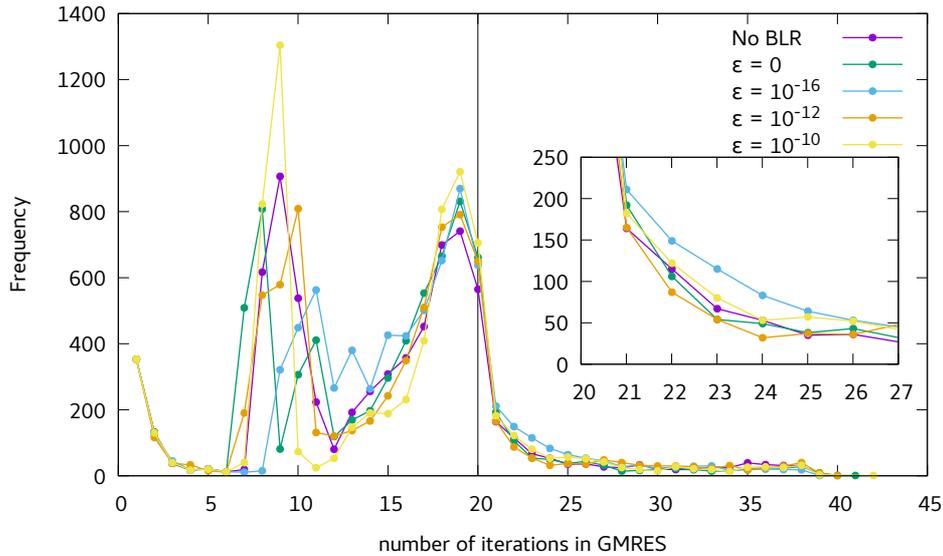}
	\caption{Histogram of GMRes iteration numbers for a tolerance of $10^{-6}$}
	\label{fig_spi_histo_gmres}
\end{figure}

In the first set of simulations with a GMRes tolerance of $10^{-6}$, the number of factorisations varied somewhat unexpectedly, as can be seen in Tab.~ \ref{tab_data_spi_runs}. Indeed, the number of factorisations rose substantially for $\epsilon = 10^{-16}$, largely exceeding the number of factorisations for $\epsilon = 10^{-12}$ and $\epsilon = 10^{-10}$. Indeed, the frequency (number of occurences) for each occuring number of GMRes iterations shown in Fig.~\ref{fig_spi_histo_gmres} reveals an increased number of GMRes iterations between $21$ and $26$ for $\epsilon = 10^{-16}$. This in turns leads to a higher average time per iteration (Tab.~\ref{tab_data_spi_runs}).

\begin{figure}[ht!]
	\includegraphics[scale=1]{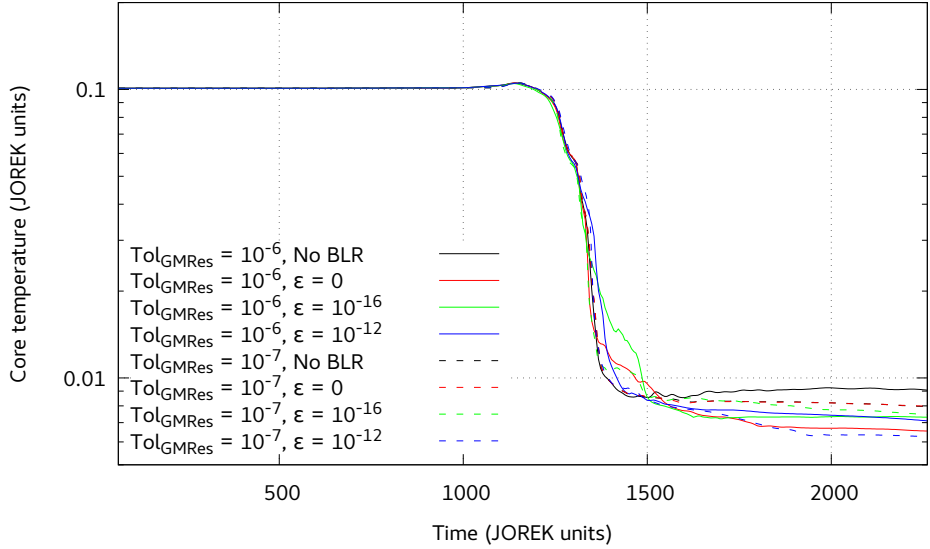}
	\caption{Core temperature evolution for GMRes tolerances of $10^{-6}$ and $10^{-7}$.}
	\label{fig_spi_core_temp}
\end{figure}

This discrepancy can be explained by the fact that a GMRes tolerance of $10^{-6}$ is too high, possibly leading to different physical results. Note that in this simulation, most factorisations take place during the physically violent thermal quench of the plasma core. Even small variations in the duration or intensity of this process can lead to the observed differences. Indeed, a longer thermal quench is observed for $\epsilon = 10^{-16}$, as shown in Fig.~\ref{fig_spi_core_temp} (at $t \sim 1300-1500$). Note that this does not directly explain why the number of factorisations  increases for $\epsilon = 10^{-16}$, but it does make this fact less surprising. The reduction of the relative differences in the number of factorisations at a higher GMRes tolerance value of $10^{-7}$ (see next section) seems to confirm that the original tolerance of $10^{-6}$ is too high.

At all observed converging values of $\epsilon$, the runtime is substantially increased, as shown by the average runtime per iteration in Tab.~\ref{tab_data_spi_runs}. This is only partly due to longer and more numerous factorisations, as the biggest contributing factor to the increased runtime is the increased runtime in GMRes, which originates from an increased runtime per solve, not from an increased number of iterations, as can be seen from Tab.~\ref{tab_data_spi_runs}. At the highest converging value of $\epsilon = 10^{-10}$, this leads to a $\sim32\%$ runtime increase for a memory gain of $\sim24\%$.

\subsubsection{Runtime performance for a GMRes tolerance of $10^{-7}$}

In the case of a GMRes tolerance of $10^{-7}$, the number of factorisations seems to be roughly constant for all values of $\epsilon$, and it is surprisingly slightly reduced compared to the simulations without BLR compression. The latter is probably again due to small differences in the physical results, as the former indicates that the accuracy of the preconditioner is determined here by the linearity of the system. Indeed, the average number of GMRes iterations and the number of factorisations does not increase for large values of $\epsilon$ up to $10^{-12}$. For $\epsilon = 10^{-8}$, this does not hold anymore, as the GMRes solver did not reach convergence in this case.

The constancy of the number of factorisations in this case means that the time per iteration is now set by the time for factorisation, solve and GMRes, all of which incur increased runtimes from (de-)compression. However, these are partially offset by the use of larger values of $\epsilon$. Even at the high number of factorisations for this GMRes tolerance value, the main contributor to the increased runtime is the runtime in GMRes, although the increased runtime for factorisation now plays a larger role compared to the higher GMRes tolerance value. At the highest observed converging value of $\epsilon = 10^{-12}$, a memory gain of $\sim20\%$ for a $\sim30\%$ runtime increase.

The above analysis suggests that the optimal $\epsilon$ value should be found somewhere between $10^{-12}$ and $10^{-8}$, which could unfortunately not be investigated due to time constraints for this study. Logically, the optimal $\epsilon$ value should be such that the inaccuracy in the preconditioner induced by the compression accuracy is similar to that inherent in our assumption of decoupled harmonics in the preconditioner. This should ensure that the time for factorisation, solve and GMRes is decreased without incurring too great an increase in the preconditioner's inaccuracy. 

Moreover, it might even be worth going beyond this limit by increasing the maximal number of GMRes iterations when using BLR compression, allowing for still larger values of $\epsilon$ which further reduce memory consumption and possibly offset the runtime increase caused by the larger number of GMRes iterations.

\newpage
\section{PaStiX version 6.x}
The new version 6.0 of the \verb|PaStiX| sparse matrix solver is still in development, but seems to be the way forward for \verb|JOREK|, as it brings new features such as Block-Low-Rank compression~\footnote{For the tests shown here, a development version equivalent to release 6.0.2 with some additional corrections was used. The hash key of the respective commit in the git repository is 17f18ce6e87de504580c27d52e5672311c413d21.}. It is however not yet MPI-parallelised, which will be implemented in version 6.1, so its current use in \verb|JOREK| is restricted to cases with one MPI task per \verb|PaStiX| instance (i.e. one task per toroidal harmonic). Apart from this, the solver is already fully functional in \verb|JOREK|, as an updated interface has been implemented in \verb|mod_poiss.f90| (equilibrium), \verb|solve_pastix_all.f90| (direct LU factorisation of entire matrix, generally used in axisymmetric runs), \verb|solve_mat_n.f90| and \verb|gmres_precondition.f90| (LU factorisation as preconditioner followed by GMRes, generally used for multiple harmonics).

The implementation can handle multiple degrees of freedom to make analysis and factorisation more efficient (switched on through the flag \verb|USE_BLOCK| in \verb|JOREK|), although the underlying matrix structure and the analysis results currently have to be expanded during the analysis phase, as the thereafter invoked calls do not yet support multiple degrees of freedom. As this expansion is as of now only implemented in the analysis phase, the analysis is currently being repeated for every time step when using \verb|PaStiX 6| with multiple degrees of freedom in \verb|JOREK|\footnote{This overhead can be avoided in the future.}. Once the \verb|PaStiX| developers have remedied to these problems, the expansion and repeated analysis should be removed in the \verb|JOREK| implementation, giving a further boost to the runtime performance.

\subsection{Benchmarking}

In the following, we present small benchmark tests of \verb|PaStiX 6| with the previously generally used \verb|PaStiX| 5.2.1 ("Release 4492"). We again use a peeling-ballooning scenario (\verb|inxflow|): where the simulation is run with 2 toroidal harmonics ($n=0,6$) in the phase of linear growth (Sec.~\ref{sec_pastix6_ntor3}).

The time evolution is computed with and without the \verb|USE_BLOCK| feature, which also the evaluation of its usefulness in different versions of \verb|PaStiX|. Furthermore, in Sec.~\ref{sec_pastix6_openmp_scan}, the scaling in the number of OpenMP threads is checked for the different \verb|PaStiX| versions by running 10 time steps in the linear growth phase with different numbers of OpenMP threads. Finally, 10 time steps are again rerun in the linear growth phase for various spatial resolutions in Sec.~\ref{sec_pastix6_resolution_scan} to check how the new \verb|PaStiX| version scales with the problem size. The basic setup has a spatial resolution of \verb|(n_flux,n_tht) = (32,40)| and the number of OpenMP threads is set to $2$.

\subsubsection{Basic simulation}
\label{sec_pastix6_ntor3}

The simulation is run for 90 time steps in the linear growth phase to obtain a first idea of the typical runtime performance but also to later assess if the OpenMP and resolution scans which are run over only 10 time steps yield the correct results. The runtimes measured in this first benchmark are given in Tab.~\ref{tab_benchmark_basic}. They are all averaged except the analysis for single-dof which is performed only once. Both single and multiple degrees of freedom (dof) are investigated (\verb|USE_BLOCK| feature).

\begin{table}[h!]
\centering
\begin{tabular}{l | c c c c}
 \multirow{2}{*}{Runtime (in s)} & \multicolumn{2}{c}{single dof} & \multicolumn{2}{c}{multiple dofs} \\
  & PaStiX 5.2.1 & PaStiX 6.0.2 & Release 5.2.1 & PaStiX 6.0.2 \\[1pt]\hline
 Analysis             & $26.1$ & $4.9$  & $0.14$  & $0.47$\\
 Factorisation        & $10.9$ & $9.5$  & $7.81$  & $7.77$  \\
 Solve                & $0.20$ & $0.15$ & $0.17$  & $0.15$  \\
 GMRes                & $1.96$ & $1.56$ & $1.79$  & $1.55$  \\
\end{tabular}
\caption{Basic benchmark of PaStiX 6}
\label{tab_benchmark_basic}
\end{table}

Considering a single degree of freedom, the time for analysis is subtantially reduced in \verb|PaStiX 6|, by a factor of $\sim 5$. However, the analysis phase only has to be repeated once per simulation (restart), such that this gain is appreciated but should not heavily influence the total runtime. 

The reason why the use of multiple degrees of freedom seems to lead to a greater speed-up in the older \verb|PaStiX| version is that the time listed under Analysis also includes the conversion of the matrix to an input usable by \verb|PaStiX|. This conversion takes slightly longer for \verb|PaStiX 6| because the latter necessitates an additional new sparse matrix structure, whereas the matrix was directly passed to \verb|PaStiX| beforehand. This additional overhead could be reduced in the future by directly using the new sparse matrix structure in \verb|JOREK|'s \verb|distribute_harmonics| routine.

The time for factorisation is also reduced in this simple case for the new \verb|PaStiX| version. The difference shrinks when multiple degrees of freedom are used, such that the times for factorisation are very similar here. Indeed, multiple degrees of freedom are not yet implemented in the factorisation part of \verb|PaStiX 6|, such that the speed-up is reduced to the contribution from a more performant analysis.

Finally, the runtimes for solution and for GMRes are reduced in the new \verb|PaStiX| version, by approximately $25\%$ when a single degree of freedom is assumed. Here, the use of multiple degrees of freedom does not lead to a speed-up for \verb|PaStiX 6|, as it is not yet implemented in the solve part.

\subsubsection{OpenMP thread scan}
\label{sec_pastix6_openmp_scan}

\begin{figure}[ht!]
	\includegraphics[scale=1]{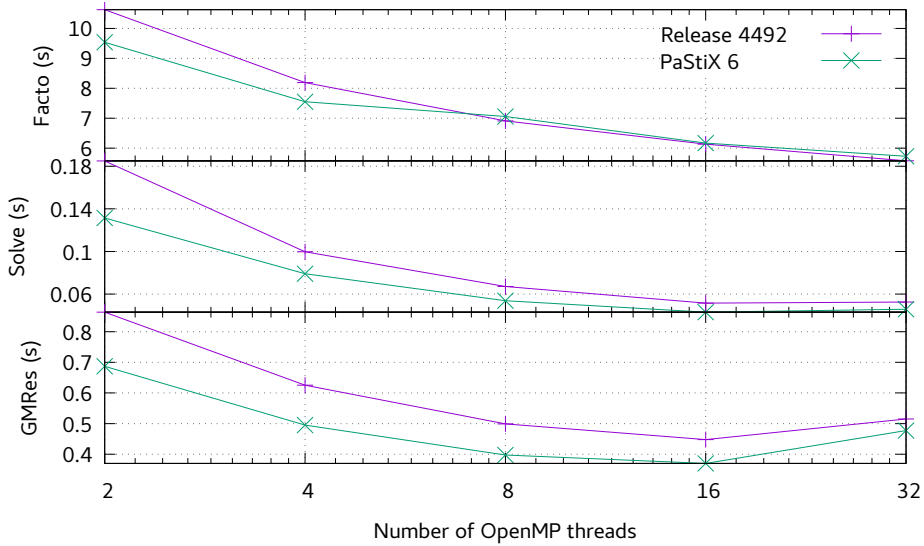}
	\caption{Runtimes in OpenMP scan for the single-dof case}
	\label{fig_benchmark_openmp}
\end{figure}

In this scan, 10 time steps were performed during the linear growth phase and the factorisation was forced to take place at every time step. The number of OpenMP threads was varied in factors of $2$ from $2$ to $32$. The resulting runtimes for the factorisation and solve steps, as well as the GMRes solver, are shown in Fig.~\ref{fig_benchmark_openmp}. The analysis step was left out as it did not vary depending on the number of OpenMP threads, so the result from the previous basic benchmark stands (old version: $\sim 25$ s, new version: $\sim 5$ s).

\begin{figure}[ht!]
	\includegraphics[scale=1]{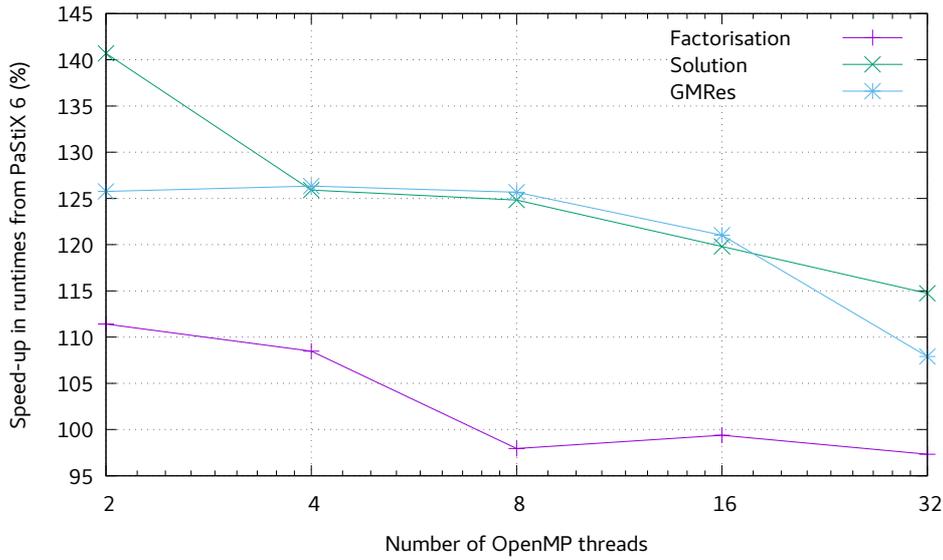}
	\caption{Speedup from PaStiX 6 in the OpenMP scan}
	\label{fig_benchmark_openmp_speedup}
\end{figure}

Although the scaling in the number of OpenMP threads seems to be worse for the factorisation step in the new \verb|PaStiX| version, the difference is minimal even at $32$ threads, where the runtimes for factorisation are basically the same between the two \verb|PaStiX| versions. The difference also becomes smaller for the solve and GMRes steps, but the new version of \verb|PaStiX| was still always faster for these compared to the old version. This can also be seen in Fig.~\ref{fig_benchmark_openmp_speedup}, where the speed-up in \verb|PaStiX 6| has been computed by dividing the runtime of the older \verb|PaStiX| version by that of \verb|PaStiX 6|. The analysis phase speed-up is $\sim 5$ for all OpenMP configurations.

\subsubsection{Resolution scan}
\label{sec_pastix6_resolution_scan}

The same 10 time steps were now performed at resolutions between \verb|(n_flux, n_tht)| = (16, 20) and (128, 160) in 6 steps of approximate size $(\sqrt{2}, \sqrt{2})$. Only $2$ OpenMP threads were used, and the factorisation was again forced to be repeated at every time step.

\begin{figure}[ht!]
	\includegraphics[scale=1]{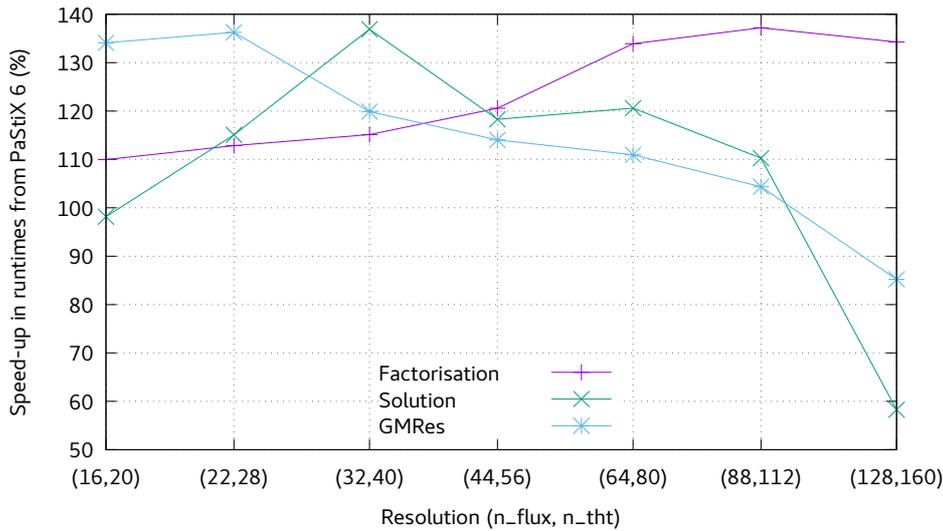}
	\caption{Speedup from PaStiX 6 in the resolution scan}
	\label{fig_benchmark_resol_speedup}
\end{figure}

The speed-up in this scan can be seen in Fig.~\ref{fig_benchmark_resol_speedup}, showing that \verb|PaStiX 6| leads to a speed-up for all phases (Factorisation, Solution, GMRes) and all resolutions, with the notable exception of the solution and GMRes phases at the very highest resolutions. The analysis phase was not included in Fig.~\ref{fig_benchmark_resol_speedup}, as it stays constant around $\sim 500\%$ for all resolutions.

The speed-up in the factorisation is most encouraging as it does not plummet at larger resolutions, which is positive considering how significant this part of the solver is (e.g. runtime for factorisation at the highest resolution employed here, \verb|PaStiX| Release 4492: $631$ s, \verb|PaStiX 6|: $470$ s). The solve and GMRes phases show a more mixed picture, as the speed-up seems to be lost when going to very high resolutions. Whether this trend continues at even higher resolutions remains to be investigated, and this trend could of course change during the further development of \verb|PaStiX 6|. Moreover, these phases are generally less critical for the total runtime (e.g. runtime for solution phase at the highest resolution employed here, \verb|PaStiX| Release 4492: $6.93$ s, \verb|PaStiX 6|: $11.9$ s). This prompts the need for a future benchmarking of \verb|PaStiX 6| on a realistic \verb|JOREK| simulation, where the effect on the total runtime could be meaningfully investigated.

\subsection{Preliminary results from BLR compression with PaStiX 6}
The new BLR compression feature in \verb|PaStiX 6| could not yet be thoroughly tested, as some convergence problems occured when using the memory-optimal settings. The following results are derived from a resolution scan with the same setup as in Sec.~\ref{sec_MUMPS_resol} but using the \verb|PaStiX 6| solver with the "just-in-time" setting, which optimises runtime gains instead of memory gains \cite{pichon:hal-01824275}.

The memory consumption, obtained from a \verb|PaStiX 6| diagnostic, is listed in Tab.~\ref{tab_memory_pastix6}. Note that the numbers may not all be accurate to the last digit, and can not necessarily be directly compared to those given in the MUMPS resolution scan. Nevertheless, they indicate that very good compression can be attained, even for small (albeit non-zero) values of $\epsilon$. This also seems to indicate that the accuracy of the BLR solver here cannot be directly compared to that from the \verb|MUMPS| solver, possibly due to an additional internal scaling in \verb|MUMPS|.

\begin{table}[!ht]
\begin{tabular}{c|c c c c c c}%
\verb|n_flux,n_tht| &  No BLR &  $\epsilon=0$ &  $\epsilon=10^{-16}$ &  $\epsilon=10^{-12}$ &  $\epsilon=10^{-8}$ &  $\epsilon=10^{-4}$ \\[1pt]\hline
$16,20$ & $  477$ & $ 476 $ & $  473$ & $  460$ & $  385$ & $  249$\\
$22,28$ & $  1070$ & $  1060$ & $  1045$ & $  968$ & $  770$ & $  470$\\
$32,40$ & $  2579$ & $  2549$ & $  2393$ & $  2112$ & $  1681$ & $  982$\\
$44,56$ & $  5570$ & $  5520$ & $  4910$ & $  4273$ & $  3398$ & $  1907$\\
$64,80$ & $  13320$ & $  13120$ & $  10970$ & $  9410$ & $  7410$ & $  4020$\\
$88,112$ & $  27770$ & $  27460$ & $  21620$ & $  18470$ & $  14600$ & $  8020$\\
$128,160$ & $  65100$ & $  64300$ & $  47600$ & $  40680$ & $  31950$ & $  17350$
\end{tabular}
\caption{Memory consumption in the PaStiX 6 resolution scan in Megabytes}
\label{tab_memory_pastix6}
\end{table}

\begin{figure}
	\includegraphics[scale=1]{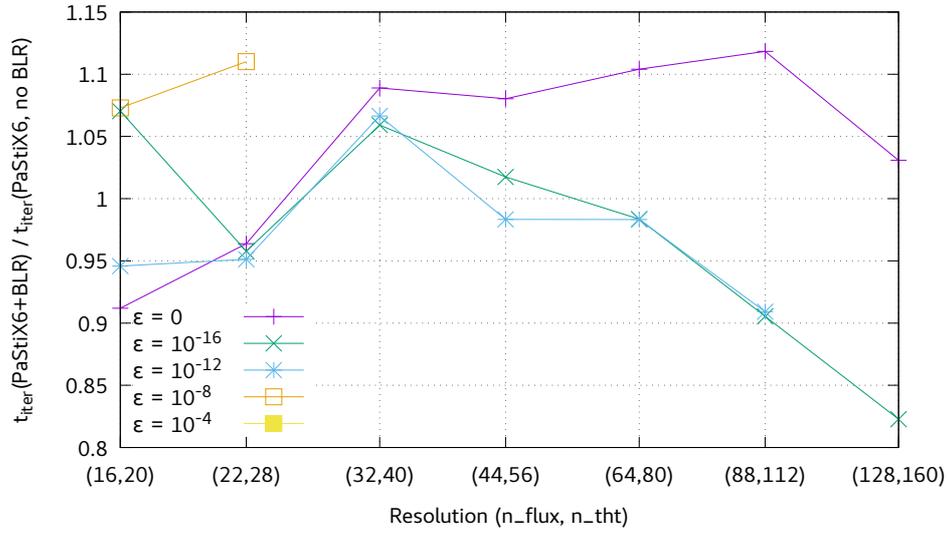}
	\caption{Average runtime per time step}
	\label{fig_resol_runtime_iter_pastix6}
\end{figure}

The speed-up in the average time per iteration in Fig.~\ref{fig_resol_runtime_iter_pastix6} confirms that the $\epsilon$ values cannot be directly compared to those of the \verb|MUMPS| solver, as many more simulations could not reach convergence here (all runs with $\epsilon = 10^{-4}$ and almost all with $\epsilon = 10^{-8}$)\footnote{Note, that version 6.0.2 of PaStiX also had a bug, which will be resolved in 6.0.3, which might partially explain this behaviour}. However, it seems easier to obtain a speed-up here, as many simulations with  $\epsilon = 10^{-12}, 10^{-16}$ demonstrate, especially at high resolutions.

\begin{figure}
	\includegraphics[scale=1]{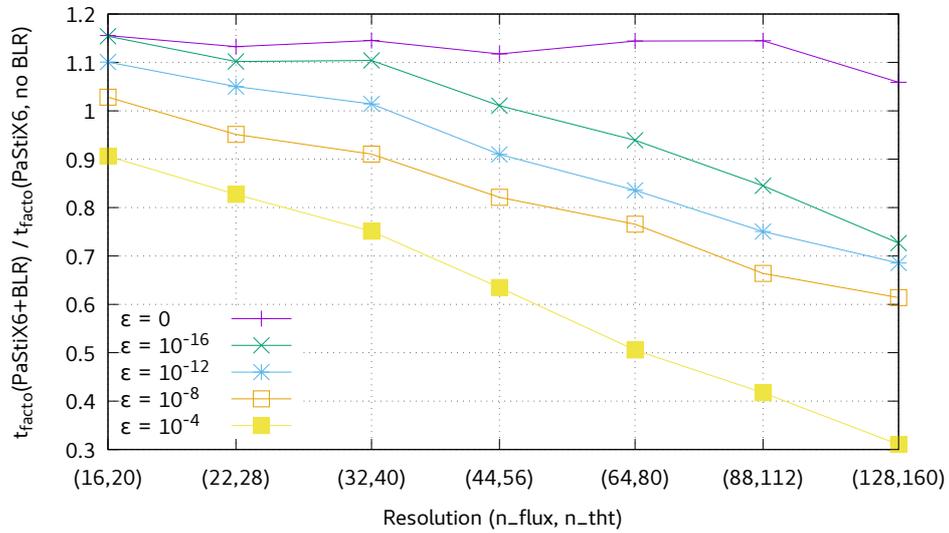}
	\caption{Runtime for factorisation step}
	\label{fig_resol_runtime_facto_pastix6}
\end{figure}

This speed-up can mostly be traced back to the speed-up in the factorisation step, shown in Fig.~\ref{fig_resol_runtime_facto_pastix6}. Note that the factorisation is forced to be repeated every time step in this scan, such that the factorisation will be the main contribution to the total time per iteration. This is not necessarily the case in a typical JOREK simulation, as Sec.~\ref{sec_spi} demonstrated.

\begin{figure}
	\includegraphics[scale=1]{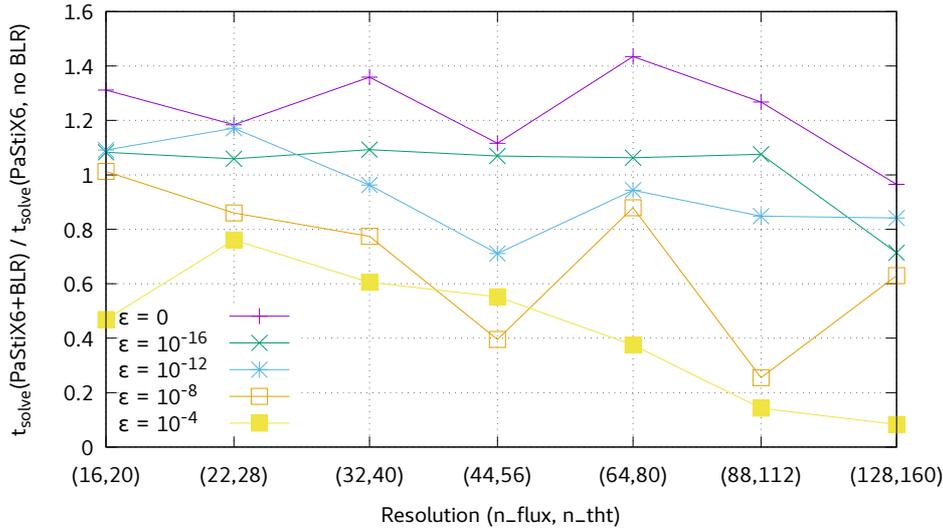}
	\caption{Runtime for solution step}
	\label{fig_resol_runtime_solve_pastix6}
\end{figure}

\begin{figure}
	\includegraphics[scale=1]{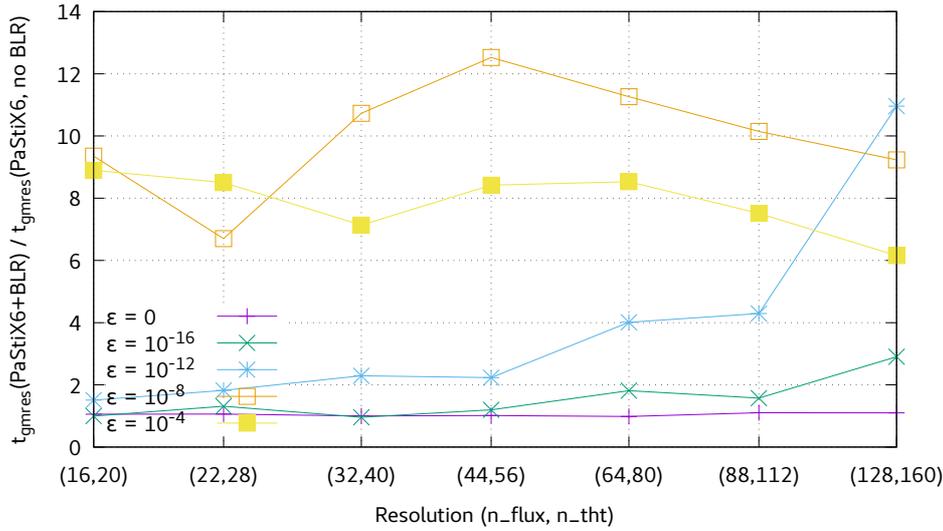}
	\caption{Runtime for GMRes}
	\label{fig_resol_runtime_gmres_pastix6}
\end{figure}

It is thus instructive to investigate the speed-up in the solve and GMRes steps, shown in Figs.~\ref{fig_resol_runtime_solve_pastix6} and \ref{fig_resol_runtime_gmres_pastix6}. The speed-up or slow-down of the solve phase depends on the value of $\epsilon$ employed, a higher resolution only mildly increases the speed-up for a given $\epsilon$. In comparison, the GMRes runtimes seem to always be longer, as a result of the worse preconditioner. At high resolutions however, the fact that the time for solve is reduced with BLR suggests that the runtime for GMRes could also be reduced in a realistic \verb|JOREK| simulation. Indeed, the number of GMRes iterations in Tab.~\ref{tab_data_spi_runs} showed that for small enough values of $\epsilon$ the accuracy of the preconditioner is primarily determined by the nonlinearity of the system, not by the BLR accuracy. In other words, the increased runtime for GMRes observed in Fig.~\ref{fig_resol_runtime_gmres_pastix6} merely reflects the highly increased number of iterations in GMRes, which were not observed in the realistic simulation (Sec.~\ref{sec_spi}) when using small enough values of $\epsilon$.

The results in this section are preliminary, as many aspects of BLR compression in \verb|PaStiX 6| remain to be investigated. However, they are already very encouraging as the memory consumption is greatly reduced and the factorisation step seems to enjoy a subtantial speed-up. Whether this speed-up can rival the slow-down in the GMRes phase in a realistic \verb|JOREK| simulation remains to be investigated.

\section{Conclusions}

Interfaces in the JOREK MHD code have been updated for MUMPS and PaStiX in order to test block low rank compression offered by these solver libraries. First tests show promising trends. Further tests are necessary to fully evaluate the benefit in production simulations.

\section*{Acknowledgements}
First and foremost, the authors would like to thank Mathieu Faverge and the rest of the \verb|PaStiX 6| development team for their quick and helpful responses to the diverse problems we encountered during the implementation of the new \verb|PaStiX 6| interface in \verb|JOREK|.
The authors would also like to thank Guido Huijsmans for his instructive comments on the use of Block-Low-Rank compression in \verb|JOREK|.

\bibliography{report_BIB}
\bibliographystyle{ieeetr}

\end{document}